\documentclass[a4paper,11pt]{article}
\usepackage[dvips]{graphicx}
\usepackage{amssymb}
\usepackage{amsmath}

\newcommand{\jmp} {J. Math. Phys. }

\begin{document}

\title{Casimir energy for a Regular Polygon with Dirichlet Boundaries}
\author{V.K.Oikonomou\thanks{
voiko@physics.auth.gr}\\
Technological Education Institute of Serres, \\
Department of Informatics and Communications 62124 Serres, Greece\\
and\\
Department of Theoretical Physics Aristotle University of Thessaloniki,\\
Thessaloniki 541 24 Greece} \maketitle

\begin{abstract}
We study the Casimir energy of a scalar field for a regular
polygon with N sides. The scalar field obeys Dirichlet boundary
conditions at the perimeter of the polygon. The polygon
eigenvalues $\lambda_N$ are expressed in terms of the Dirichlet
circle eigenvalues $\lambda_C$ as an expansion in $\frac{1}{N}$ of
the form,
$\lambda_N=\lambda_C{\,}(1+\frac{4{\,}\zeta(2)}{N^2}+\frac{4{\,}\zeta(3)}{N^3}+\frac{28{\,}\zeta(4)}{N^4}+\ldots
{\,})$. A comparison follows between the Casimir energy on the
polygon with $N=4$ found with our method and the Casimir energy of
the scalar field on a square. We generalize the result to spaces
of the form $R^d\times P_N$, with $P_N$ a N-polygon. By the same
token, we find the electric field energy for a ''cylinder'' of
infinite length with polygonal section. With the method we use and
in view of the results, it stands to reason to assume that the
Casimir energy of $D$-balls has the same sign with the Casimir
energy of regular shapes homeomorphic to the $D$-ball. We sum up
and discuss our results at the end of the article.
\end{abstract}

\section*{Introduction}

More than fifty years have passed since the article of H. Casimir
\cite{Casimir} in which the force between conducting parallel
plates was studied. The article was initially not so well known
but the experiments where great advocates of its theoretical
predictions and lead up to the establishment of the article making
it a foundation for new research. The years that followed up to
now where full of theoretical and experimental (for example see
\cite{lamoreaux}) research towards the study of the so-called
Casimir energy and Casimir force for various geometrical
configurations \cite{Bordagreview}. Also the technological
applications of the Casimir force are of great importance, for
example in nanotubes, nano-devices in general and in
microelectronic engineering \cite{micro}. Indeed it is obvious
that the attractive or repulsive nature of the Casimir force can
lead to the instability or even destruction of such a
micro-device. In this manner, the study of various geometrical and
material configurations will lead us to have control over the
Casimir force, stabilizing the last by using appropriate
geometrical structures.

\noindent The Casimir effect finds its explanation in the quantum
structure of the vacuum. Indeed due to the fluctuations of the
electromagnetic quantum field the parallel plates interact (in
fact attracted towards each other). In addition it was of profound
importance that the boundaries do alter the quantum field boundary
conditions and as a result the plates interact. Indeed it was
realized that the geometry of the boundaries have a strong effect
in the Casimir energy and Casimir force, rendering the last
repulsive or attractive. In Casimir's work the force was
attractive and some years later Boyer \cite{Boyer} studied the
conducting sphere case, where it was found that the force is
repulsive. These advisements where generalized to include other
quantum fields such as fermions, bosons and other scalar fields
making the Casimir effect study a widespread and necessary
ingredient of many theoretical physics subjects such as string
theory, cosmology \cite{elizaldegenika,odintsov} e.t.c. For a
concise treatment on these issues see for example
\cite{Bordagreview,elizaldegenika,odintsov,kirstenkaialloi,oikonomoureview}
and references therein.

\noindent By now it is obvious that the boundary conditions affect
drastically the Casimir force for all the aforementioned quantum
fields. The most used boundary conditions are the Dirichlet and
Neumann boundary conditions. However, these boundary conditions
have no direct generalization in the case of fermion fields and in
general for fields with spin$\neq 0$ \cite{ambjorn}. In those
cases, the bag boundary conditions are used which were introduced
to provide a solution to confinement \cite{bag}.

\noindent To our knowledge, up to date very few studies related to
the Casimir energy for polygons where done. Some of them included
the computation of the Casimir energy for hyperbolic polygons
\cite{turk}, also for tetrahedra \cite{dowker} and for triangles
\cite{iaponas}. The main difficulty was maybe the absence from the
mathematical literature of the eigenvalue computation for regular
polygons with Dirichlet boundaries. In this article we present one
solution to such a problem, borrowed from the recent mathematical
literature. This solution is found utilizing the calculus of
moving surfaces \cite{gilbert1,gilbert2} which in our case is a
perturbation of the circle with Dirichlet boundary conditions
(from now on Dirichlet circle). The regular polygons are
homeomorphic to the circle. This gives us the opportunity to model
this by employing a kind of perturbation, as we mentioned. We
shall see that the eigenvalues are expressed as a series in $1/N$,
where $N$ stands for the number of sides of the regular polygon.
When $N\rightarrow \infty$ we recover the Dirichlet circle's
eigenvalues as is expected, since the method is perturbative and
does not rely to collapsing coordinates that parameterize the
regular polygon when $N\rightarrow \infty$. As an application we
shall compare our result for $N=4$ with the Casimir energy of a
scalar filed in a Dirichlet square. A numerical and an analytic
study of both will help us to make the comparison. A
generalization of the Casimir energy for spaces of the form
$R^d\times P_N$ follows ($P_N$ is the $N$-polygon).

\noindent The interest in studying the Casimir energy in such
polygonal configurations stems from various applications, both
theoretical and experimental and also from conceptual interest.
The last is a good motivation because the curvature singularities
that corners introduce are usually hard to handle and so being
able to find solutions to Dirichlet problems in such
configurations is quite a challenge. The theoretical applications
are focused mainly in the fact that such spaces arise in the $1+2$
dimensional theory of gravity \cite{dowker}. Also these spaces can
be considered as triangulation approximations to smooth manifolds
(in a way the last will be used in our article). Both
experimentally and theoretically, polygonal configurations arise
in superconducting regular polygons. We shall briefly discuss
these issues later on in this article.

\section*{Laplace Eigenvalues on Regular Polygons and Casimir Energy}

Consider the Laplace equation for a scalar field, solved for a
circle, with the field obeying Dirichlet boundary conditions on
the circle, namely:
\begin{equation}\label{eq1}
-\Delta \Phi(x,y)=\lambda \Phi(x,y),
\end{equation}
with $\Delta$, $\lambda$ and $\Phi(x,y)$, the Laplacian, the
eigenvalues of the Laplacian and the scalar field respectively. It
is much more convenient to express the equation and solutions in
polar coordinates,
\begin{equation}\label{eq1}
-\Delta \Phi(r,\phi)=\lambda \Phi(r,\phi).
\end{equation}
Let $\lambda_N$ and $\lambda_C$ denote the eigenvalues of the
problem corresponding to the Dirichlet regular polygon and to the
Dirichlet circle respectively. The eigenvalues of the polygon
$\lambda_N$ were studied in \cite{gilbert1,gilbert2} by employing
the calculus of moving surfaces. According to this method the
polygon is treated as a non-smooth perturbation of the circle. The
problem was first studied by Hadamard in \cite{hadamard} and the
solution is given as an integral over the mesh, that is, the area
between the circle and the inscribed regular polygon:
\begin{equation}\label{hadamard}
\frac{\lambda_N}{\lambda_C}=\int \int_{C-P_N} |\mathrm{grad}{\,}
u_C|\mathrm{d}x\mathrm{d}y.
\end{equation}
Hadamard proved that the first correction to (\ref{hadamard}) is
of order $\sim 1/N^2$. The authors of \cite{gilbert1,gilbert2}
improved the solution of Hadamard, giving three more terms in the
perturbation series, as we shall shortly see.

\noindent Without getting into much detail, let us shortly
describe here the philosophy of the method. It is based on the
calculus of moving surfaces. The key point is to find the shape
derivative, the derivative of $\lambda$ as the boundary shape
changes. The calculus of moving surfaces is based on the rate at
which the boundary moves normal to itself. The last is fully
described by the fundamental theorem for a moving surface $S$
bounding a region $\Omega$,
\begin{equation}\label{theorem}
\frac{\mathrm{d}}{\mathrm{d}\tau}\int_{\Omega}F\mathrm{d}\Omega=\int_{\Omega}\frac{\partial
F}{\partial\tau}{\,}\mathrm{d}\Omega+\int_S C{\,}F{\,}\mathrm{d}S.
\end{equation}
In the above equation $C$ stands for the velocity of the boundary
and $F$ is an invariant. Additionally to theorem (\ref{theorem})
there exists the theorem,
\begin{equation}\label{theorem1}
\frac{\mathrm{d}}{\mathrm{d}\tau}\int_{\Omega}F\mathrm{d}S=\int_{S}\frac{\mathrm{\delta}
F}{\mathrm{\delta}\tau}{\,}\mathrm{d}S-\int_S
C{\,}B_{\alpha}^{\alpha}{\,}\mathrm{d}S.
\end{equation}
Mention that the derivative $\frac{\mathrm{\delta}
F}{\mathrm{\delta}\tau }$, is defined to act as follows,
\begin{equation}\label{derivative}
\frac{\mathrm{\delta} F}{\mathrm{\delta}\tau }=\frac{\partial
F}{\partial \tau }+C{\,}\frac{\partial F}{\partial n},
\end{equation}
while $B_{\alpha}^{\alpha}$ is the trace of the curvature tensor
$B^{\alpha}_{\beta}$. In brief, invariance on surfaces is achieved
by introducing the covariant derivative $\nabla_a$, while
invariance on moving surfaces can be achieved by using the
$\delta/\delta \tau$ derivative, which was firstly introduced by
Hadamard \cite{hadamard}. The $\delta/\delta \tau$ derivative acts
on the invariant field $F$ of the moving surface. An invariant
field of the surface can be for example, the mean curvature, the
eigenfunction $\psi$ of the Laplace eigenvalue equation for the
area $\Omega$, or the velocity of the surface boundary $C$. For
more details consult references \cite{gilbert1,gilbert2} and
references therein. We shall take the result of their method for
computing the eigenvalues.

\noindent So making use of the method of moving surfaces and
treating the polygon as a perturbation of the circle, the
eigenvalues of the polygon $\lambda_N$ are related to the
eigenvalues of the Dirichlet circle problem $\lambda_C$ according
to the following series expansion in powers of $1/N$:
\begin{equation}\label{basicrelationa}
\lambda_N=\lambda_C{\,}(1+\frac{4{\,}\zeta(2)}{N^2}+\frac{4{\,}\zeta(3)}{N^3}+\frac{28{\,}\zeta(4)}{N^4}+\ldots
{\,}),
\end{equation}
where $\zeta$ is the Riemann zeta function
\cite{gilbert1,gilbert2}. Call to mind that the eigenvalues
$\lambda_C$ are the roots $\gamma_{mn}$ of the Bessel functions of
the first kind $J_m(\gamma_{mn}{\,}r)$, and that $u_C\sim
J_m(\gamma_{mn}{\,}r)e^{im\phi}$. The method works for the radial
eigenfunctions, so the $\phi$ dependence is dropped, but this does
not affect our analysis.

\noindent Before closing this section we must mention a drawback
that such calculations have. It is the fact that there are
singularities at the corners of the polygon. It is obvious that
sharp corners introduce qualitatively new features that cannot be
easily captured by perturbative techniques. However an
infinitesimal smoothing of the corners does not alter the
eigenfunctions concentrated away from the boundaries
\cite{gilbert1}. Likewise, the moving surfaces approach yields
expressions that remain valid even for large eigenvalues (see
\cite{gilbert1} and references therein).

\subsection*{Casimir Energy}

The Casimir energy for the regular polygon is the sum of the
energy eigenvalues for that space with Dirichlet boundary
conditions, that is,
\begin{equation}\label{casimirenergy}
E_N=\frac{1}{2}\sum_{n}\sqrt{\lambda_N^{(n)}},
\end{equation}
where the sum over $n$ denotes all the modes that will be summed
in the end. Substituting relation (\ref{basicrelationa}) in
(\ref{casimirenergy}), we obtain:
\begin{equation}\label{casimirenergy1}
E_N\simeq
\frac{1}{2}\sum_{n}\sqrt{\lambda_C^{(n)}{\,}\Big{(}1+\frac{4{\,}\zeta(2)}{N^2}+\frac{4{\,}\zeta(3)}{N^3}+\frac{28{\,}\zeta(4)}{N^4}}\Big{)},
\end{equation}
Recall that the Dirichlet circle Casimir energy is equal to:
\begin{equation}\label{casimirenergy3}
E_C=\frac{1}{2}\sum_{n}\sqrt{\lambda_C^{(n)}},
\end{equation}
hence the polygon Casimir energy reads,
\begin{equation}\label{casimirenergy4}
E_N\simeq
\sqrt{\Big{(}1+\frac{4{\,}\zeta(2)}{N^2}+\frac{4{\,}\zeta(3)}{N^3}+\frac{28{\,}\zeta(4)}{N^4}}\Big{)}E_C
\end{equation}
In the large $N$ limit ($N\rightarrow \infty$), the regular
polygon Casimir energy reads,
\begin{equation}\label{largen limit}
E_N\simeq
E_C+\frac{7\pi^4}{45}\frac{1}{N}E_C+\frac{1350\pi^2-49\pi^8}{4050}\frac{1}{N^2}E_C+\ldots
\end{equation}
By looking the above dependence it is obvious that the Dirichlet
circle Casimir energy is recovered in the large $N$ limit of the
regular polygon's Casimir energy. To see how valid is the
approximation we made in relation (\ref{casimirenergy4}), let us
compare the perturbative result of (\ref{casimirenergy4}) for
$N=4$ with the Casimir energy of a square area with Dirichlet
boundary conditions on each side of the square. Take for
simplicity that the radius of the circle is 1, thus the side of
the square is $1$.
\begin{figure}[h]
\begin{center}
\includegraphics[scale=1.0]{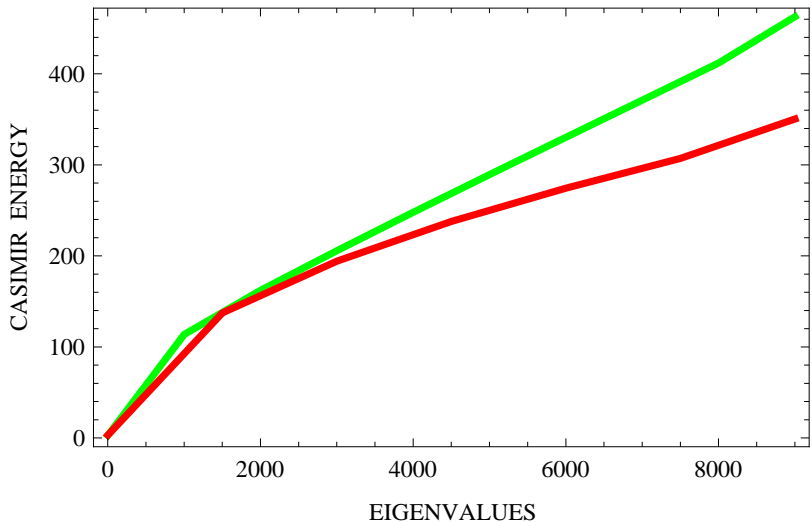}
\end{center}
\caption{The Casimir energy of the polygon (green) and of the
square (red) for $10^4$ eigenvalues} \label{casimirplot1}
\end{figure}

\begin{figure}[h]
\begin{center}
\includegraphics[scale=1.0]{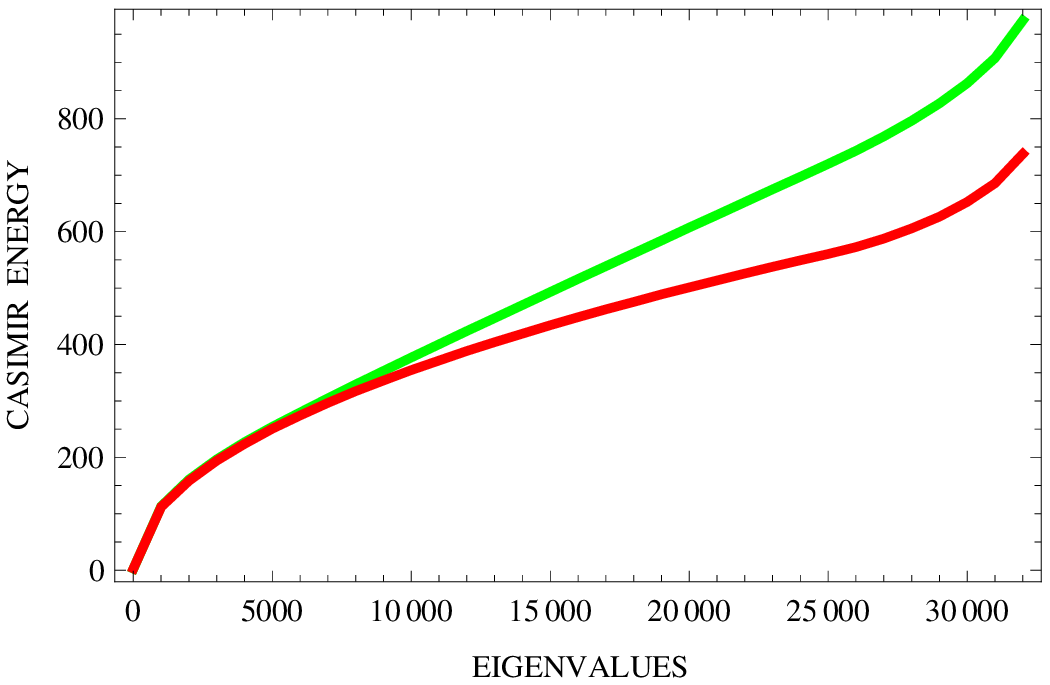}
\end{center}
\caption{The Casimir energy of the polygon (green) and of the
square (red) for $32400$ eigenvalues} \label{casimirplot2}
\end{figure}

\begin{figure}[h]
\begin{center}
\includegraphics[scale=1.0]{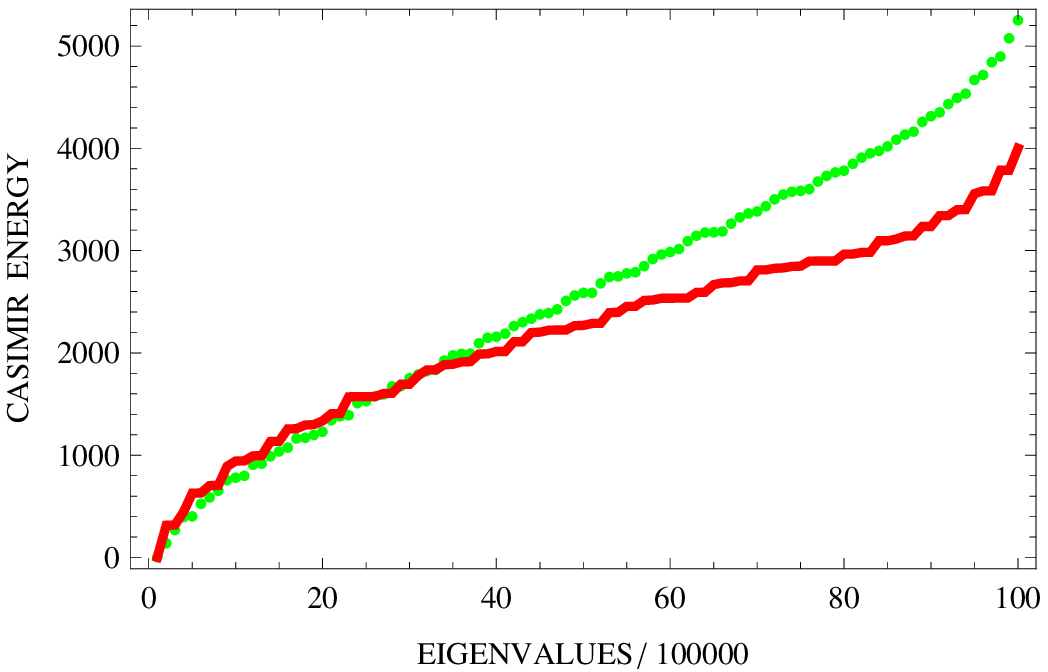}
\end{center}
\caption{The Casimir energy of the polygon (green) and of the
square (red) for $10^6$ eigenvalues} \label{casimirplot3}
\end{figure}
\noindent The eigenvalues of the Laplacian for the Dirichlet
square are equal to:
\begin{equation}\label{squareeig}
\lambda_s=\frac{\pi}{a}\sqrt{m^2+n^2},
\end{equation}
with $a=1$. The eigenvalues $\lambda_C$ of the Dirichlet unit
circle are the roots of the Bessel function of first kind,
$J_n(x_{nm})$, that is $\lambda_C=x_{nm}$. In such away, the
polygon eigenvalues for $N=4$ considering the polygon as a circle
perturbation are:
\begin{equation}\label{polygeigenu}
\lambda_{nm}\sim
\sqrt{\Big{(}1+\frac{4{\,}\zeta(2)}{N^2}+\frac{4{\,}\zeta(3)}{N^3}+\frac{28{\,}\zeta(4)}{N^4}}\Big{)}x_{nm},
\end{equation}
with $N=4$. Let us check how the eigenvalues (and accordingly the
Casimir energy) behaves as a function of $n$ and $m$. In figures
(\ref{casimirplot1}), (\ref{casimirplot2}) and
(\ref{casimirplot3}) we plot the comparison of the first $10^4$,
$32400$ and $10^6$ eigenvalues of the Dirichlet unit circle and
the corresponding square ones. We can see how close the two plots
are, proving that the approximation is quite valid. On that
account, the perturbation of the circle eigenvalues gives results
that fail to be truth for a percentage about $16.2\%$, as it is
found numerically. Let us check the asymptotic form of
eigenvalues. Remember that the asymptotic behavior of the Bessel's
function $J_n(x)$ roots is given by:
\begin{equation}\label{asym}
x_{mn}=n\pi+(m-\frac{1}{2})\frac{\pi}{2}.
\end{equation}
and thereupon the polygon eigenvalues are:
\begin{equation}\label{polygeigenu1}
\lambda_{nm}\sim
\sqrt{\Big{(}1+\frac{4{\,}\zeta(2)}{N^2}+\frac{4{\,}\zeta(3)}{N^3}+\frac{28{\,}\zeta(4)}{N^4}}\Big{)}\Big{(}n\pi+(m-\frac{1}{2})\frac{\pi}{2}\Big{)}
\end{equation}
We compare the above with these of relation (\ref{squareeig}) for
various values of $n$ and $m$. For $n\gg m$, the square
eigenvalues are equal to $\lambda_s\sim \pi n$, while the
polygon's are $\lambda_{nm}\sim n\pi
\sqrt{\Big{(}1+\frac{4{\,}\zeta(2)}{N^2}+\frac{4{\,}\zeta(3)}{N^3}+\frac{28{\,}\zeta(4)}{N^4}}\Big{)}$
which for $N=4$ is approximately,
\begin{equation}\label{polygeigenu2}
\lambda_{nm}\sim 1.26678{\,}n\pi.
\end{equation}
Along these lines, we can see that the eigenvalues are
approximately equal with a difference of $20\%$. For $m\sim n$, in
which case $\lambda_{nm}$ becomes $\lambda_{nm}\sim
1.90017{\,}n\pi$ and also $\lambda_s\sim 2\pi n$. So the
difference of the two above is approximately $5\%$. Now the only
problem arises when $m \gg n$, in which case we have,
$\lambda_s\sim \pi m$ and $\lambda_{nm}\sim
1.26678{\,}n\frac{\pi}{2}$. So the difference in these eigenvalues
approaches $57\%$. However as we saw previously, the actual
numerical difference is much smaller. In fact the numerical
calculation performed for a very large range of eigenvalues gave a
difference which approaches $16.2\%$. This result is adequate, in
view of the fact that we are working with a perturbative method.

\noindent We give now a much more elaborate calculation of the
Casimir energy for both the square and the polygon, using the zeta
function regularization
\cite{elizaldegenika,odintsov,kirstenkaialloi,oikonomoureview}. In
the case of the Bessel zeros sums, although the sum is explicit,
the analytic continuation is not easy to do. The square Casimir
energy can be done easily employing zeta regularization
\cite{elizaldegenika,odintsov,kirstenkaialloi,oikonomoureview}.

\subsection*{Casimir energy of the Dirichlet circle $B_2$}

Consider a scalar field obeying Dirichlet boundary conditions on a
circle. We shall find the Casimir energy of the scalar field. With
the same notation as above, the Casimir energy is equal to,
\begin{equation}\label{casz}
E_C=\frac{1}{2}\sum_{m,n}x_{m,n}.
\end{equation}
The corresponding zeta function is,
\begin{equation}\label{zeta1s}
\zeta (s)=\sum_{m,n} \lambda _{m,n}^{s}(a)-\sum_{m,n}
\bar{\lambda}_{m,n}^{s}(\infty).
\end{equation}
In the above, $\lambda_{m,n}(a)$ and $\bar{\lambda}_{m,n}(\infty)$
are the scalar field's eigenvalues with Dirichlet boundaries and
when the boundary is sent to infinity respectively. The Casimir
energy can be written as,
\begin{equation}\label{casz1}
E_C=\frac{1}{2}\zeta (s=-1).
\end{equation}
Following the techniques developed in
\cite{elizaldecontour,nesterenko}, we find that,
\begin{equation}\label{spectralzeta}
\zeta (-1)=\frac{1}{\pi} \lim_{s\rightarrow 0}\sum_{i=0}^{3}Z_i(s)
\end{equation}
with $Z_0$ to stand for,
\begin{equation}\label{z0}
Z_0(s)=-(1-s)\int_0^{\infty}\mathrm{d}y{\,}y^{-s}\Big{[}\ln
(2y\mathrm{I}_0(y)\mathrm{K}_0(y))-\frac{t^2}{8}(1-6t^2+5t^4)\Big{]},
\end{equation}
$Z_1$ being equal to,
\begin{equation}\label{z1}
Z_1(s)=\frac{1-s}{2}\zeta (s-1)\Gamma
(-\frac{1-s}{2})\sum_{m=1}^{\infty}\frac{\Gamma
(m-\frac{1-s}{2})}{m\Gamma (m)},
\end{equation}
$Z_2$ equal to,
\begin{equation}\label{z2}
Z_2(s)=\frac{1}{2}\big{(}\zeta
(s+1)\frac{1}{2}\big{)}\Gamma(\frac{1+s}{2})(-1+3(1+s)-\frac{5}{8}(3+s)(1+s))
\end{equation}
and finally $Z_3$,
\begin{align}\label{z3}
Z_3(s)=&\frac{1}{32}\zeta (s+3)\Gamma (\frac{3-s}{2})(-13\Gamma
(\frac{3+s}{2})+142\Gamma (\frac{5+s}{2})-177\Gamma
(\frac{7+s}{2}) \\& \notag+\frac{113}{2}\Gamma
(\frac{9+s}{2})-\frac{113}{24}\Gamma (\frac{11+s}{2})).
\end{align}
In the above equation, $\mathrm{I}_0$ and $\mathrm{K}_0$ are the
modified Bessel functions of first and second kind, and
$t=1/\sqrt{1+y^2}$. Relations (\ref{z0}), (\ref{z1}), (\ref{z2})
and (\ref{z3}) for $s=0$ after numerical calculation and
integration become (note that as $s\rightarrow 0$, $\zeta(1+s)\sim
\frac{1}{s}+\gamma +\ldots $),
\begin{align}\label{z01}
Z_0(0)&=\pi(0.02815-\frac{1}{128}),{\,}{\,}{\,}Z_1(0)=-\frac{\pi}{12},{\,}{\,}{\,}
\\ & \notag Z_2(0)=\frac{\pi}{64}(\frac{1}{s}+\gamma
)+\frac{\pi}{128},{\,}{\,}{\,}Z_3(0)=-0.13679\frac{\pi}{64}\zeta
(3).
\end{align}
Adding the above contributions we obtain,
\begin{equation}\label{zetaref}
\zeta(-1)=\frac{1}{a}\big{(}0.047189-\frac{1}{64{\,}s}\big{)},
\end{equation}
and finally the Casimir energy is equal to,
\begin{equation}\label{casdf}
E_C=\frac{1}{a}\big{(}0.023595-\frac{1}{128{\,}s}\big{)}.
\end{equation}
The drawback of relation (\ref{zetaref}) in company with
(\ref{casdf}) is the existence of a pole. This fact makes someone
to doubt about the validity of the above expressions. However it
is the only way to obtain a clue of what the regularized Casimir
energy looks like. But in the end the result is not to be trusted
\cite{elizaldecontour}. Like so, the numerical value of the
Casimir energy $E_C$ for the unit circle is,
$E_C=0.023595-\frac{1}{128{\,}s}$. Call to mind that the polygon
Casimir energy is equal to,
\begin{equation}\label{casimireny4}
E_N\simeq
\sqrt{\Big{(}1+\frac{4{\,}\zeta(2)}{N^2}+\frac{4{\,}\zeta(3)}{N^3}+\frac{28{\,}\zeta(4)}{N^4}}\Big{)}E_C,
\end{equation}
hence we obtain that, $E_N\simeq
0.029769-\frac{1.266783}{128{\,}s}$.

\subsection*{Casimir energy of the Dirichlet square}

In the case of the Casimir energy of a scalar field in a Dirichlet
square, things are more easy and accurate compared to the
Dirichlet circle case. Indeed the zeta function for the Dirichlet
square is \cite{elizaldegenika}:
\begin{equation}\label{ff}
\zeta(s)=\sum_{m,n}\Big{(}\frac{\pi^2m^2}{a^2}+\frac{\pi^2n^2}{a^2}\Big{)}^{-s}.
\end{equation}
By virtue of the following asymptotic expansion,
\begin{align}\label{gone1}
\zeta_{EH}(s;p)&=\frac{1}{2}\sum_{n=-\infty}^{\infty
'}\big{(}n^2+p\big{)}^{-s}
\\& \notag =-\frac{p^{-s}}{2}+\frac{\sqrt{\pi}{\,}{\,}\Gamma(s-\frac{1}{2})}{2\Gamma(s)}{\,}p^{-s+\frac{1}{2}}+\frac{2\pi^s{\,}p^{-s+\frac{1}{2}}}{\Gamma(s)}\sum_{n=1}^{\infty}n^{s-\frac{1}{2}}K_{s-\frac{1}{2}}\big{(}2\pi
n\sqrt{p}\big{)},
\end{align}
relation (\ref{ff}) can be written as:
\begin{align}\label{squareexp}
\zeta(s)&=-\frac{1}{2}\Big{(}\frac{a}{\pi}\Big{)}^{2s}\zeta
(2s)+\frac{a}{2\sqrt{\pi}}\Big{(}\frac{a}{\pi}\Big{)}^{2s-1}\frac{\Gamma(s-1/2)}{\Gamma(s)}\zeta(2s-1)
\\ &
\notag+\frac{2}{\Gamma(s)}\Big{(}\frac{a^2}{\pi}\Big{)}^{s}\sum_{m,n}\Big{(}\frac{m}{n}\Big{)}^{s-1/2}\mathrm{K}_{s-1/2}(2\pi
m n),
\end{align}
with $\mathrm{K}_{\nu}$ the modified Bessel function. Thereupon,
the numerical value of the Casimir energy $E_S$ for $a=1$ is
$E_S=0.0415358$. Let us gather the results at this point. We take
the finite part of the Dirichlet circle Casimir energy and we have
that the difference in the Casimir energies of the polygon and of
the square is approximately $28\%$, which is quite bigger than the
one we found numerically earlier in this article (see below
relation (\ref{polygeigenu2})). Of course we must note that the
pole in the Dirichlet circle Casimir energy does not allow us to
be sure of the $\frac{E_S-E_N}{E_S}$ difference. Consequently we
must realize that the approximation we made in the beginning of
this article,
\begin{equation}\label{basicrelation21}
\lambda_N=\lambda_C{\,}(1+\frac{4{\,}\zeta(2)}{N^2}+\frac{4{\,}\zeta(3)}{N^3}+\frac{28{\,}\zeta(4)}{N^4}+\ldots
{\,}),
\end{equation}
clearly holds with an $16.2\%$ accuracy for a wide range of modes,
but a deeper investigation is needed to get a clear and accurate
result on this.

\section*{The Casimir energy on $R^{{\,}D}\times P_N$ and on a Cylinder}

Let us generalize the results of the previous section by
considering a scalar field in a spacetime with topology
$R^{{\,}D}\times P_N$, where $P_N$ denotes the regular polygon.
The scalar field is assumed to obey Dirichlet boundary conditions
on the regular polygon perimeter. We shall make use of the zeta
function regularization method in order to obtain regularized
results for the Casimir energy. We can write the Casimir energy
for the total spacetime in the following form:
\begin{align}\label{vlad}
\mathcal{E}_N=  \frac{1}{4\pi^2}\int
\mathrm{d}^{D-1}p\sum_{m,n}\Big{[}\sum_{k=1}^{D-1}p_k^2+\lambda_{mn}\Big{]}^{-s}.
\end{align}
Remember that $\lambda_{mn}$ stand for the regular polygon
eigenvalues. In addition we shall take $s=-1/2$ in the end in
order to recover the Casimir energy. Upon integrating over the
continuous dimensions utilizing the following,
\begin{equation}\label{feynman}
\int
\mathrm{d}k^{D-1}\frac{1}{(k^2+A)^s}=\pi^{\frac{D-1}{2}}\frac{\Gamma(s-\frac{D-1}{2})}{\Gamma(s)}\frac{1}{A^{s-\frac{D-1}{2}}}
\end{equation}
relation (\ref{vlad}) is reformed into,
\begin{align}\label{pordoulis}
\mathcal{E}_N=\frac{1}{4}\pi^{\frac{D-5}{2}}\frac{\Gamma(s-\frac{D-1}{2})}{\Gamma(s)}
\sum_{m,n}\big{(}\lambda_{mn}\big{)}^{\frac{D-1}{2}-s}.
\end{align}
Bring to mind that,
\begin{equation}\label{basicrelation1}
\lambda_{mn}=x_{mn}{\,}(1+\frac{4{\,}\zeta(2)}{N^2}+\frac{4{\,}\zeta(3)}{N^3}+\frac{28{\,}\zeta(4)}{N^4}+\ldots
{\,}),
\end{equation}
where $x_{mn}$ stand for the Dirichlet circle eigenvalues. In
virtue of the above two relations we obtain that,
\begin{align}\label{pordoulis1}
\mathcal{E}_N\simeq
\frac{1}{4}\pi^{\frac{D-5}{2}}\frac{\Gamma(s-\frac{D-1}{2})}{\Gamma(s)}
\Big{(}1+\frac{4{\,}\zeta(2)}{N^2}+\frac{4{\,}\zeta(3)}{N^3}+\frac{28{\,}\zeta(4)}{N^4}\Big{)}^{\frac{D-1}{2}-s}\sum_{m,n}\big{(}x_{mn}\big{)}^{\frac{D-1}{2}-s}.
\end{align}
The expression,
\begin{align}\label{porlis1}
\frac{1}{4}\pi^{\frac{D-5}{2}}\frac{\Gamma(s-\frac{D-1}{2})}{\Gamma(s)}\sum_{m,n}\big{(}x_{mn}\big{)}^{\frac{D-1}{2}-s}.
\end{align}
is equal to the Casimir energy $\mathcal{E}_C$ of a scalar field
in a spacetime with topology $R^{D}\times B_2$, with $B_2$ the
Dirichlet disc. Thereupon, relation (\ref{pordoulis1}) can be
written:
\begin{equation}\label{nuo}
\mathcal{E}_N\simeq
\Big{(}1+\frac{4{\,}\zeta(2)}{N^2}+\frac{4{\,}\zeta(3)}{N^3}+\frac{28{\,}\zeta(4)}{N^4}\Big{)}^{\frac{D-1}{2}-s}
\mathcal{E}_C
\end{equation}
Taking the limit $N\rightarrow \infty$, relation (\ref{nuo}) gives
the expected results, with the polygon Casimir energy becoming
equal to the Dirichlet circle. The analysis of the validity of
relation (\ref{nuo}) can be done following the same steps as we
did in the previous section.

\subsection*{Case of Infinite cylinder with a regular polygon as section}

Consider a deformed cylinder of height $a$ and with a regular
polygon as a section. We can easily generalize the calculation we
made in the previous subsection. Indeed the Casimir energy for the
deformed cylinder, $\mathcal{E}_{Ncyl}$, can be written as:
\begin{align}\label{vlad21}
\mathcal{E}_{Ncyl}=  \frac{1}{4\pi^2}\int
\sum_{k=1}^{\infty}\sum_{m,n}\Big{[}(\frac{k\pi}{a})^2+\lambda_{mn}\Big{]}^{\frac{1}{2}}.
\end{align}
Exerting the asymptotic expansion,
\begin{align}\label{gone}
\zeta_{EH}(s;p)&=\frac{1}{2}\sum_{n=-\infty}^{\infty
'}\big{(}n^2+p\big{)}^{-s}
\\& \notag =-\frac{p^{-s}}{2}+\frac{\sqrt{\pi}{\,}{\,}\Gamma(s-\frac{1}{2})}{2\Gamma(s)}{\,}p^{-s+\frac{1}{2}}+\frac{2\pi^s{\,}p^{-s+\frac{1}{2}}}{\Gamma(s)}\sum_{n=1}^{\infty}n^{s-\frac{1}{2}}K_{s-\frac{1}{2}}\big{(}2\pi
n\sqrt{p}\big{)}
\end{align}
and by subtracting the Casimir energy for a deformed cylinder with
infinite length we obtain:
\begin{align}\label{vlfg1}
\mathcal{E}_{Ncyl}=
-\frac{1}{2\pi}\sum_{m,n}\sum_{k=1}^{\infty}\frac{\sqrt{\lambda_{mn}}K_1(2{\,}k\sqrt{\lambda_{mn}}{\,}a)}{n}.
\end{align}
This relation is perfectly suited for finding asymptotic limits.
For example when the argument of the Bessel function is large, the
Bessel function can be very well approximated by the following
relation,
\begin{equation}\label{besselappr}
K_{\nu}(z)=\sqrt{\frac{\pi}{2z}}e^{-z}\Big{(}1+\frac{\nu-1}{8z}+....\Big{)}.
\end{equation}
In our case $\nu =1$ so the Casimir energy for very large
 $a$ is equal to:
\begin{align}\label{vlasym}
\mathcal{E}_{Ncyl}\sim
-\frac{1}{2\pi}\sum_{m,n}\sum_{k=1}^{\infty}\frac{\sqrt{\pi
\lambda_{mn}}e^{-k\sqrt{\lambda_{mn}}{\,}a}}{n\sqrt{{2k\sqrt{\lambda_{mn}}{\,}a}}}.
\end{align}
Bring to mind that,
\begin{equation}\label{basicrelation21}
\lambda_{mn}=x_{mn}{\,}(1+\frac{4{\,}\zeta(2)}{N^2}+\frac{4{\,}\zeta(3)}{N^3}+\frac{28{\,}\zeta(4)}{N^4}+\ldots
{\,}),
\end{equation}
hence relation (\ref{vlasym}) is converted to,
\begin{align}\label{vlasym12}
\mathcal{E}_{Ncyl}\sim -\frac{1}{2\pi}\frac{\sqrt{\pi
x_{mn}}e^{-\sqrt{\mathcal{N}{\,}x_{mn}}{\,}a}}{n\sqrt{{2\sqrt{\mathcal{N}{\,}x_{mn}}{\,}a}}}.
\end{align}
with
$\mathcal{N}=1+\frac{4{\,}\zeta(2)}{N^2}+\frac{4{\,}\zeta(3)}{N^3}+\frac{28{\,}\zeta(4)}{N^4}$
and $x_{mn}$ the Dirichlet ball eigenvalues for the scalar field.
The above relation holds for the lowest eigenvalues $x_{mn}$.
Before closing let us note that the TM modes of the
electromagnetic field, inside a perfectly conducting resonator of
length $a$ with a regular polygon section, are exactly the
eigenvalues of relation (\ref{vlad21}), that is,
\begin{equation}\label{electric}
\omega_{m,n}=\Big{[}(\frac{k\pi}{a})^2+\lambda_{mn}\Big{]}^{\frac{1}{2}}.
\end{equation}
The calculation of the Casimir energy is straightforward and is
equal to (\ref{vlad21}). Consequently the relations proved in this
subsection apply to the electric field modes in a deformed
cylindrical perfect conducting resonator.

\section*{Discussion and conclusions}

In this article we studied the problem of having a scalar field
confined inside a regular polygon and obeying Dirichlet boundary
conditions at the perimeter of the polygon. Our objective was to
calculate the Casimir energy of the scalar field, which reduces in
finding the eigenvalues of the scalar field inside the polygon. We
used a result from the mathematical literature
\cite{gilbert1,gilbert2} that connects the eigenvalues of the
regular polygon to those of the Dirichlet circle (also known as
Dirichlet ball $B_2$ a subcase of the $D$-dimensional ball
\cite{elizaldecontour}). The method used in
\cite{gilbert1,gilbert2} involved the calculus of moving surfaces
which is actually a method that treats the regular polygon as a
perturbation of the circle. The relation we used is,
\begin{equation}\label{basicrelation}
\lambda_N=\lambda_C{\,}(1+\frac{4{\,}\zeta(2)}{N^2}+\frac{4{\,}\zeta(3)}{N^3}+\frac{28{\,}\zeta(4)}{N^4}+\ldots
{\,}),
\end{equation}
with $\lambda_{N}$ and $\lambda_{C}$ the polygon and circle
eigenvalues respectively. We have calculated the results both
numerically and analytically. The numerical calculations where
done for a large number of eigenvalues, $10^4$, $32400$ and
$10^6$. We found an $16.2\%$ difference between the two
eigenvalues, with the polygon eigenvalues being larger. Also the
analytical method gave a $28\%$ difference. However we should call
in question the last result, because the circle eigenvalues
contain a pole. We must note that the numerical calculation showed
us that the difference in the two eigenvalues remains
approximately $16\%$ for a wide range of eigenvalues (we tried for
the first $10^6$).

\noindent In this manner, making use of the method of moving
surfaces enables us to gain insight on how the circle eigenvalues
change under the deformation of the circle to a homeomorphic to it
regular shape. This is very useful on it's own because it can give
us a hint on how the Casimir energy tends to behave as the
geometry, shape and maybe topology of an initial configuration
changes. Of course the boundary conditions must remain the same
for both the shapes under study. By common consent, the Casimir
energy strongly depends on the geometry, topology and boundary
conditions of the configuration for which it is computed. However
we don't have a general rule on how the sign of the Casimir energy
depends on these. Additionally we don't have a rule on how the
Casimir energy changes as one of the geometry or topology changes.
In reality, all our theoretical results are restricted because all
Casimir energy calculations are performed for shape preserving
geometries. On that account, we don't have an answer on how the
Casimir energy behaves when the boundary deforms (the shape, not
boundary conditions). The method of moving surfaces gives us a
hint to a (maybe) general rule. In virtue of the results presented
in this article we could say that the Casimir energy of
homeomorphic (of course not difeomorphic) configurations have the
same sign (for Dirichlet boundaries). This holds for two
dimensional (surfaces) and three dimensional configurations.
Additionally we have strong evidence that the homeomorphism
includes symmetry in the deformation procedure, that is all shapes
during the deformation must be symmetric in some way. This
includes both the initial and the final shape (it is like a
symmetric (reversible) deformation process). For a shape with
corners, symmetry means that the shape must be regular. This
clearly holds for the Dirichlet circle and all known regular
polygons, like the triangle \cite{iaponas} and the square
\cite{Bordagreview}. Also it is known that the sphere and the cube
have the same sign in the Casimir energy, but the rectangular
cavity can be negative (but the rectangular polyhedron is not
regular). Additionally the regular pyramid \cite{gamotourkos} has
positive Casimir energy, just like the sphere. Clearly there is
much work to be done in this direction before we can be sure that
the above holds with certainty. For sure we have a good hint on
how homeomorphic deformations affect the Casimir energy of an
initial configuration. It is like shape perturbation theory.

\noindent A question arises quite naturally. Can the opposite
hold, that is, if we know a regular polygons Casimir energy
(eigenvalues), can we find the circle Casimir energy
(eigenvalues)? This could be very useful, if it holds, because we
know analytically for example the square's Casimir energy and so
we maybe have a way to know the circle's Casimir energy without
the (unwanted) pole singularity.

\noindent The interest in polygonal surfaces arises both
theoretically but also experimentally. Theoretical applications
can be found in $2+1$ dimensional theory of gravity, in
microelectronics and in superconductivity, as we already said. In
reference to superconductivity, there are polygonal
superconducting constructions. Let us discuss this further. With
the progress made in microfabrication techniques it is possible to
investigate microscopic superconducting samples with sizes smaller
than the coherence length and penetration depth. It is known by
now that the boundary geometry has a strong effect on the
nucleation of superconductivity in the samples. The boundary
conditions that are usually imposed for the order parameter $\psi$
are:
\begin{equation}\label{supercond}
\Big{(}-i\hbar \nabla -\frac{2e}{c}A\Big{)}\psi\Big{|}_n=0,
\end{equation}
with $A$ the vector potential corresponding to the magnetic field.
The presence of a vector potential seriously complicates the
solution of the Landau-Ginzburg equation for arbitrarily shaped
samples. However in the $A_n=0$ (i.e. the normal component along
the boundary line is zero) gauge the simplification is obvious.
The remaining problem reduces to one with Neumann boundary
conditions (which is similar to the Dirichlet problem we solved).
This gauge for the vector field is applicable when the screening
effects of the supercurrents can be neglected. It has been applied
and by now experimentally tested in triangles and squares. Also
the $A=0$ gauge is a common choice for configurations such as
infinite slabs, disks and semi-planes with a wedge (see
\cite{supercond} for the $A=0$ gauge examples. Useful are the
discussions of
\cite{oikonomoumajorana,oikonomoupistons,oikonomoureview} and
references therein). The generalization of superconducting samples
to geometries such as a regular polygons was done in
\cite{supercond}. The regular polygon configurations are of great
importance since these appear frequently in Josephson junctions,
where different superconducting elements with characteristic size
in the $\mu\mathrm{m}$ range meet each other. New vortex patterns
could be found in superconducting regular polygons, displaying an
anti-vortex in the center of the polygon for some values of the
applied magnetic field flux (for the connection of
vortex-anti-vortex, Majorana modes and Casimir energy see the
discussion of \cite{oikonomoumajorana} and references therein).
These can be probed by scanning tunnelling microscopy.

\noindent The calculus of moving surfaces is a very useful tool
for calculating eigenfrequencies of dynamically deformed spaces.
It would be very interesting trying to apply these techniques to
the dynamical Casimir effect. Some applications appearing in the
literature  are the shape optimization of electron bubbles
\cite{gilbert3}, the calculation of the gravitational potential
for deformations of spherical geometries \cite{gilbert4}. For an
alternative perturbative approach to ours see the recent article
\cite{signal}. In addition the analytic calculation of Casimir
energies and the associated spectral zeta functions in spaces with
spherical boundaries, involves uniform asymptotic expansions of
Bessel functions products \cite{elizaldecontour,nesterenko}. The
lack of such expansions in problems with boundary geometry
different from spherical is an obstacle in making asymptotic
expansions and hence raises difficulties towards the calculation
of the corresponding Casimir energy. It is probable that with the
aid of moving surfaces calculus we may have a solution for the
Casimir energies of deformed boundaries as a function of the
Casimir energy corresponding to spherical boundaries. A simple
application of the calculus of moving surfaces is the computation
of the eigenvalues for a slowly uniformly inflating disk. Let
$\lambda_{\Delta R}$ and $\lambda_R$ denote the eigenvalues of the
final disk with radius $r+\Delta R$ and of the initial disk with
radius $r$ (we shall use $r=1$). The two eigenvalues are related
to each other by,
\begin{equation}\label{eiginflat}
\lambda_{\Delta R}^{(n)}=\lambda_R{(n)}(1-2\Delta R+3\Delta
R^2+\ldots),
\end{equation}
The Casimir energy of the inflated disk in terms of the other one
is given by:
\begin{equation}\label{casimirinflated}
E_{\Delta R}\sim \sqrt{1-2\Delta R+3\Delta R^2}{\,}E_{R}.
\end{equation}
With $\Delta R$ taking values for which the perturbation still
holds, we have $\Delta R \gg \Delta R^2$. In this view, relation
(\ref{casimirinflated}) evolves into,
\begin{equation}\label{casimirinflated11}
E_{\Delta R}\sim \sqrt{1-2\Delta R}{\,}E_{R}.
\end{equation}
Note that when the radius decreases, $\Delta R<0$, thereupon, the
Casimir energy $E_{\Delta R}$ increases, while when the radius
increases, then $\Delta R>0$, in which case the Casimir energy
decreases. Both cases are compatible with what physical intuition
lead us to think. Indeed as the radius decreases we expect the
Casimir energy to increase. The calculus of moving surfaces can
give results for much complex evolution of curves
\cite{gilbert1,gilbert2,gilbert3}.

\section*{Acknowledgments}

V. Oikonomou is indebted to Prof. G. Leontaris for the hospitality
at the University of Ioannina, where part of this research was
conducted.


\begin{thebibliography}{99}


\bibitem{Casimir} H. Casimir, Proc. Kon. Nederl. Akad. Wet. 51 793
(1948)

\bibitem{lamoreaux} S. K. Lamoreaux, Phys. Rev. Lett. 78, 5 (1997) ;A.O. Sushkov, W. J. Kim,
D. A. R. Dalvit, S. K. Lamoreaux, arXiv:1011.5219

\bibitem{Bordagreview} M. Bordag, U. Mohideen, V. M. Mostepanenko,
Phys. Rep. 353, 1 (2001); Michael Bordag, Galina Leonidovna
Klimchitskaya, Umar Mohideen, Vladimir Mikhaylovich Mostepanenko,
''Advances in the Casimir Effect'', International Series of
Monographs on Physics, Oxford University Press (2009)


\bibitem{micro} F. M. Serry, D. Walliser, G. J. MacLay, Journal of Microelectromechanical Systems 4, 193 (1995)

\bibitem{Boyer} T. H. Boyer, Phys. Rev. 174 (1968) 1764

\bibitem{elizaldegenika} E. Elizalde, J. Phys. A41, 304040 (2008); E. Elizalde, J. Phys. A39, 6299
(2006); S. D. Odintsov, I. L. Buchbinder, Fortsch. Phys. 37, 225
(1989); K. Milton, Phys. Rev. D22, 1444 (1980); K. Milton, Phys.
Rev. D22, 1441 (1980); E. Elizalde, ``Ten physical applications of
spectral zeta functions", Springer (1995); E. Elizalde,
S.~D.~Odintsov, A. Romeo, A. A. Bytsenko, S. Zerbini, ``Zeta
regularization techniques and applications", World Scientific
(1994); E. Elizalde J. Phys. A 41, 304040 (2008)

\bibitem{odintsov} I. L. Buchbinder, S. D. Odintsov, Sov. Phys. J. 27, 554 (1984); E. Elizalde, S. D.
Odintsov, A. Romeo, J. Math. Phys. 37, 1128 (1996); E. Elizalde,
S. Nojiri, S. D. Odintsov, S. Ogushi, Phys. Rev. D67, 063515
(2003); I. L. Buchbinder and S. D. Odintsov, Int. J. Mod. Phys.
A4, 4337 (1989); Fortshrt. Phys. 37, 225 (1989); S. D. Odintsov,
Sov. Phys. J. 31, 695 (1988) E. Elizalde, S. D. Odintsov and S.
Leseduarte, Phys. Rev D49, 5551 (1994); I. Brevik, K. Milton, S.
Nojiri and S. D. Odintsov, Nucl. Phys. B599, 305 (2001); S. D.
Odintsov, Sov. Phys. J. 27, 554 (1984)


\bibitem{kirstenkaialloi} K. Kirsten, J. Phys. A26, 2421 (1993); K. Kirsten, J. Phys. A25, 6297 (1992); K. Kirsten, J. Phys. A24, 3281 (1991)
; Klaus Kirsten, Spectral Functions in Mathematics and Physics,
Chapman Hall/CRC (2001); K. Kirsten, J. Math. Phys. 35, 459; K.
Kirsten, J. Math. Phys. 32, 3008-3014 (1991); I. L. Buchbinder, S.
D. Odintsov Sov. Phys. J. 26, 359 (1983); S. D. Odintsov, Mod.
Phys. Lett. A3, 1391 (1988); S. D. Odintsov, Phys. Lett. B306, 233
(1993);  E. Elizalde, S. D. Odintsov, A. Romeo, Phys. Rev. D54,
4152 (1996); E. Elizalde, S. Nojiri, S. D. Odintsov, S. Ogushi,
Phys. Rev. D67, 063515 (2003); E. Elizalde, A. Romeo, Int. J. Mod.
Phys. A 7, 7365 (1992)

\bibitem{oikonomoureview} V. K. Oikonomou, Rev. Math.  Phys. 21, 615 (2009); V. K. Oikonomou, arXiv: 0905.4928; V. K. Oikonomou, J.
Phys. A40, 5725 (2007)

\bibitem{ambjorn} J. Ambjorn, S. Wolfram, Ann. Phys. 147, 1 (1983)

\bibitem{bag} A. Chodos, R. L. Jaffe, K. Johnson, C. B. Thorn, V. F. Weisskopf, Phys. Rev. D9, 3471 (1974)

\bibitem{turk} H. Ahmedov, J. Phys. A 40, 10611 (2007)

\bibitem{dowker} J. S. Dowker, \jmp 28, 33 (1987)

\bibitem{iaponas} Norio Inui, J. Phys. Soc. Jpn. 76, 114002 (2007)

\bibitem{gilbert1} P. Grinfeld, G. Strang, Computers and Mathematics with
Applications, 48, 1121 (2004)

\bibitem{gilbert2} P. Grinfeld, Boundary Perturbations of Laplace
Eigenvalues. Applications to Polygons and Electron Bubbles
Department of Mathematics, MIT, 2003

\bibitem{hadamard} J. Hadamard,  Memoires presentes a l'Academie
des Sciences, vol. 33, 23 (1908)

\bibitem{elizaldecontour} G. Cognola, E. Elizalde, K. Kirsten, J. Phys. A34, 7311
(2001); E. Elizalde, M. Bordag, K. Kirsten, J. Phys. A31, 1743
(1998); K. A. Milton, L. L. DeRaad, Jr., J. S. Schwinger, Annals
Phys. 115, 388 (1978)

\bibitem{nesterenko} V. V. Nesterenko, I. G. Pirozhenko, J. Math. Phys. 41, 4521
(2000)

\bibitem{gamotourkos} H. Ahmedov, I.H. Duru, J. Math. Phys. 46, 022304
(2005); H. Ahmedov, I.H. Duru, J. Math. Phys. 46, 022303 (2005);
H. Ahmedov, I.H. Duru, J. Math. Phys. 45, 965 (2004)


\bibitem{supercond}L. F. Chibotaru, A. Ceulemans, G. Teniers, V. Bruyndoncx, V. V.
Moshchalkov, Eur. Phys. J. B 27, 341 (2002)

\bibitem{oikonomoumajorana} V. K. Oikonomou, N.D. Tracas,
arXiv:0912.4825

\bibitem{oikonomoupistons}V. K. Oikonomou, Mod. Phys. Lett. A24,  2405 (2009)

\bibitem{gilbert3} P. Grinfeld, Numerical Functional Analysis and Optimization,
30, 689 (2009)

\bibitem{gilbert4} P. Grinfeld, J. Wisdom, Quart. Appl. Math., Vol. 64,
No 2, 229

\bibitem{signal} A. R. Kitson, A. I. Signal, arXiv: 1011.4055

\end{thebibliography}
\end{document}